\documentclass[12pt,a4paper,final]{iopart}

\usepackage{iopams}
\usepackage{graphicx}
\usepackage[breaklinks=true,colorlinks=true,linkcolor=blue,urlcolor=blue,citecolor=blue]{hyperref}
\usepackage{subcaption}
\usepackage{cite}
\usepackage{color}
\usepackage[utf8]{inputenc}

\begin{document}

\title[Coexistence of continuous variable QKD with intense DWDM classical channels ]{Coexistence of continuous variable QKD with intense DWDM classical channels }

\author{Rupesh Kumar, Hao Qin, Romain All\'eaume }
\address{Telecom ParisTech, LTCI CNRS, 46 rue Barrault 75013 Paris, France}
\ead{rupesh.kumarps@telecom-paristech.fr}

\begin{abstract}
We demonstrate experimentally the feasibility of continuous
variable quantum key distribution (CV-QKD) in
dense-wavelength-division multiplexing networks (DWDM), where QKD
will typically have to coexist with several co-propagating
(forward or backward) C-band classical channels whose launch power
is around 0dBm.
%We first characterize experimentally the excess
% noise induced by classical channel on a balanced homodyne
% detection.  This relates to measuring the noise associated with
% Raman scattering, that constitutes the main source of noise on
%CV-QKD in a DWDM environment. After a careful calibration of the
% parameters of our CV-QKD experiment,
%implementing the gaussian-modulated coherent state
% protocol with a system excess noise $0.03$,
% we have measured the
% total excess noise when CV-QKD is multiplexed with an intense
% classical channel, as a function of launch power and classical
% channel wavelength.
We have conducted experimental tests of the coexistence of CV-QKD
multiplexed with an intense classical channel, for different input
powers and different DWDM wavelengths. Over a 25km fiber, a CV-QKD
operated over the  1530.12nm channel can tolerate the noise arising
from up to 11.5dBm classical channel at 1550.12nm in forward
direction  (9.7dBm in backward). A positive key rate (0.49kb/s)
can be obtained at 75km with classical channel power of
respectively -3dBm and -9dBm in forward and backward. Based on
these measurements, we have also simulated the excess noise and
optimized channel allocation for the integration of CV-QKD in some
access networks.  We have, for example, shown that CV-QKD could
coexist with 5 pairs of
 channels (with nominal input powers: 2dBm  forward and 1dBm
  backward) over a 25km WDM-PON network.
 %, while the coexistence could be
%extended to 75 km, with 2 pairs of channels, if their launch power
% is reduced to -10 dBm.
The obtained results demonstrate the outstanding
capacity of CV-QKD to coexist with classical signals of realistic
intensity in optical networks.
% It also illustrates how this
%coexistence capability can boost the integration of QKD into
% existing optical networks.
\end{abstract}

\maketitle

\section{Introduction}

Quantum Key Distribution (QKD) \cite{BB84:ProcIEE84} is the only
cryptographic protocol that allows to distribute a secure key between two distant parties with a
security guarantee that holds against a computationally unbounded adversary.
QKD has attracted considerable interest over the past thirty years, with theoretical breakthrough regarding our understanding of QKD security as well as tremendous progress on the experimental side \cite{sca:rmp09}. Targeting real-world applications, fully functional prototypes as well as commercial systems have now been available from more than 10 years and deployed in test  and production environments \cite{jkl:optexp, Peev:njp09, swissquantum}. Conversely, QKD has become the most advanced quantum communication technology, with a fast-paced increase in the demonstrated performances: a thousand-fold increase of the key rates and distance limitations that have improved from about 50 km to more than 100 km \cite{NoteKeyRate}.

Most of the effort on QKD system
design and  experimental demonstrations have however so far been realized
on dark fiber \cite{Elliot:arXiv05, Peev:njp09,
Sasaki:optExp11,tnz:natphoton07,jkl:natphot13}. This
restricts the deployability of QKD to a limited number of
scenarios where the barriers associated with dark fiber availability and price can both be overcome.
On the other hand, Wavelength Division Multiplexing (WDM) allows to share a single optical
fiber to transport multiple optical channels using different wavelengths. WDM compatibility of quantum and classical communications would allow to deploy QKD on lit fiber. This would boost the  compatibility of quantum communications with existing optical infrastructures and lead to a significant improvement in terms of cost-effectiveness and addressable market for QKD.

However, coexistence with intense classical channels raises new
challenges for QKD. The optical power used on optical classical
channels is orders of magnitude higher than for quantum
communication. Multiplexing classical and quantum signals on a
single fiber can result in very important additional noise for the
quantum communication, due to insufficient isolation or to optical
non-linear effects \cite{Chraplyvy:jlightTech90}. Coping with such
noise is in general a major problem for QKD systems and filtering
techniques are needed to improve the ratio between quantum signal
and WDM-induced noise. The implementation of this filtering can
result in additional losses and severely impact the performance of
QKD. This is in particular the case for systems that rely on
spectrally wide-band single photon detectors\cite{Subacius:apl05}.
We demonstrate in this article that Continuous Variable-QKD
(CV-QKD) systems, relying on coherent detectors - that are
spectrally selective by design - can exhibit superior tolerance to
WDM-induced noise.

Pioneering work on QKD and wavelength division multiplexing  has
been performed at the very early days of QKD research by Paul
Townsend and coworkers  \cite{Townsend:elelett97}, with  one
classical channel at 1550nm multiplexed with a quantum channel at
1300nm. This corresponds to a Coarse Wavevelength Division
Multiplexing  (CWDM) configuration, that has been studied in
several other
 works \cite{chapuran:njp09, choi:njp11}. The large spectral separation the quantum and the classical
 channels presents the advantage of reducing the amount of noise due to Raman scattering
 (that is approximately 200nm wide) onto the quantum channel. CWDM QKD integration
 configuration has however several limitations:  it can only accommodate shorter distance
  (due to the higher attenuation for QKD at 1300nm)  and the coexistence is limited to a
  small number (below 8) of classical channels due to the large inter-channel spacing in CWDM.
  One could thus use this configuration in priority in the context of  access networks,
  where it seems best suited \cite{AleksicNOC13}.

On the other hand, if QKD is to be transported over long-distance
links (beyond 50km) and in coexistence with a large number
classical channels, which is the case in core or wide-area optical
networks, then Dense Division Wavelength Multiplexing (DWDM) is
required. DWDM compatibility, i.e. the capacity to coexist with
standard optical channels, all multiplexed in the C band, with
relatively narrow channel spacing (from 1.6nm to 0.2nm), is the
focus of the present article.
 DWDM compatibility of QKD has initially been studied in the work of Peters et. al.\cite{peters:njp09}, where Raman noise was identified as the main impairment for links longer than a few km. A coexistence test of QKD with two forward-propagating classical channel and a total input power of 0.3 mW has been performed. Despite this input power below typical optical network specifications and the use of some filtering, QKD could not be operated beyond 25 km. More recently, several new DWDM compatibility experiments have been performed, with discrete variable QKD (DV-QKD) systems and more efficient filtering techniques. In \cite{eraerds:njp12}, 4 classical channels
 where multiplexed with a DV-QKD system and 50km operation was demonstrated. However,
 the input power of the classical channels was attenuated below -15dBm, to the smallest
 possible power compatible with the sensitivity limit of the optical receiver (-26dBm).
 This technique was also used in  \cite{patel:prx12} with an input power limited down to
 -18.5dBm and in addition the use of a temporal filtering
 technique developed in \cite{choi:njp11} to obtain a range of 90km.

The extended working range for DV-QKD coexistence with DWDM channels demonstrated
in \cite{eraerds:njp12, patel:prx12} is of practical interest. However,
these demonstrations have been performed with strongly attenuated classical channels,
 more that 15dB below the standard level of optical input power commonly used in existing
 optical network. This indicates the difficulty of integrating DV-QKD in optical networks
 in coexistence with standard optical power (around 0dBm). Coexistence of DV-QKD with 
 0dBm channels has however recently been demonstrated over 25 km \cite{Patel:apl14},  but it requires additional use of fine-tuned time and spectral filtering. We demonstrate in the present article that the use of CV-QKD could be advantageous in order to deploy QKD in DWDM coexistence with standard
  optical channels: we have shown that  CV-QKD can coexist with classical channels whose cumulated
  power could be as high as 11.5dBm at 25km.  We have also demonstrated QKD operation (with a key rate of 0.49kbit/s)
  at 75km in coexistence with a -3dBm channel. This stronger coexistence capability can be obtained without any
  additional filtering and could be of significant advantage in many practical situations related to QKD integration
  in standard optical networks.

%%% donner les raisons de la comparaison favorable avec DVQKD, time-bandwith product

This paper is arranged as follows. In section \textbf{II} the
general principle of  CV-QKD is given. Section \textbf{III}
analyzes the different noise sources of photons generated by
classical channels on the quantum channel in a DWDM network
context. In section \textbf{IV}, we characterize experimentally
the impact of  the main source of noise: Spontaneous Anti-Stockes
Raman Scattering. We then describe in section \textbf{V} the
experiments we have performed to test the DWDM coexistence of a
CV-QKD channel with an intense (0 - 8mW) classical DWDM channel.
The analysis of the results and a comparison with previous works
are proposed in section \textbf{VI}, along with a discussion of
the integration of CV-QKD in WDM-PON networks. Finally, concluding
remarks are given in section \textbf{VII}.

\section{Gaussian modulated coherent state protocol: protocol and parameter estimation}

The Gaussian-modulated coherent state (GMCS) protocol \cite{Grosshans:prl02} was the first proposed CV-QKD protocol that could be implemented with  coherent states (and not squeezed states) \cite{GG02}, making it a attractive candidate for practical developments. There are other CV-QKD protocols proposed based on squeezed or entangled states\cite{Garcia:prl09, Hillery:pra00, Madsen:natcom12, Ralph:pra99}. However, the security analysis of GMCS protocol is moreover firmly established, with a security proof that holds against collective \cite{R09} and coherent attacks \cite{RN06}, and that can be extended to the composable security framework \cite{AnthonyArxiv2014}.

 It is probably the CV-QKD protocol that has been most widely studied and implemented, and we will also use it in our experimental implementation, which will ease the comparison of our results on DWDM coexistence with previous results for CV-QKD implemented over dark fiber.

In the GMCS, Alice encodes information on coherent states of
light, by realizing a Gaussian modulation on both quadratures
$\{X_A, P_A\}$ with the same variance $V_A$. This signal travels
through the public quantum channel, that can be characterized, in
the context of the asymptotic security against collective attack
(which we will adopt in this article), by two parameters: the
transmission $T$ (in power) and the excess noise $\xi$ (equivalent
excess noise at the input). At reception, Bob randomly chooses to
measure one of the two quadratures, $X$ or $P$, with a homodyne
detector and publicly announce which quadrature he has measured.

Alice and Bob will then reveal part of their data to estimate the
channel parameters $T$ and $\xi$. Provided the estimated
parameters are compatible with secure key generation, which can be
learned from the security proof, then key distillation, i.e. error
correction followed by privacy amplification,  will be performed on
the non-revealed data to generate key.

By symmetry we can assume without loss of generality that Bob has measured the quadrature $X$, and thus that Alice and Bob share respectively the correlated continuous variables $(X^1_A  \cdots X^N_A)$  and $(X^1_B  \cdots X^N_B)$ in $\mathbb{R}^N$.

To perform parameter estimation, i.e. estimate $T$ and $\xi$ and then secret key rate, Alice and
Bob  reveal part of their data and can make use of Eq.1-4\cite{paul:pra12}:
\begin{eqnarray}
var(X_A) &\equiv &V_A\\
 \langle X_A X_B \rangle &=& \sqrt{\eta_{B} T} V_A \\
var(X_B) &=& \eta_{B} T V_A + N_0+\eta_{B} T \xi+
v_{ele}\\
var(X_{B_0}) &=& N_0+
v_{ele}
\end{eqnarray}

Here $\eta_B$ - the transmission efficiency of Bob and  $v_{ele}$ -
the variance of the electronic noise are calibrated values,
evaluated offline and stable during an experiment. Eq.2 and 3 are
therefore sufficient to evaluate $T$ and $\xi$, provided shot
noise variance $N_0$ is calibrated, which needs to be done by an
auxiliary measurement, corresponding to Eq.4.

 Shot noise variance   calibration is performed by shining only the local oscillator on the homodyne detector
  and measuring the variance of the homodyne detector output. However, as we shall discuss later in the text, shot noise variance cannot in general be assumed to be strictly constant, and must be sampled periodically. Shot noise calibration is performed by nullifying the signal entering the input port, for example by shutting down all sources of incoming light or by closing the signal optical port on Bob side.
Throughout the article, all value homogeneous to quadrature
variances will implicitly be expressed in shot noise units (1 shot
noise unit = 1 $N_0$).

During the experiment, and for each
transmission distance, $V_A$ is chosen in order to optimize the
 secure key rate under collective attacks \cite{R09}.
The optimal $V_A$ value is determined by measuring system excess noise
for a range of $V_A$ and then by selecting the $V_A$ value that gives rise to the
maximum key rate. In our experimental system, assuming 95$\%$ of reconciliation
efficiency, the optimal value of $V_A$ was found to be 3.5$N_0$ at 25km and 2$N_0$ for
50km and 75km.

\section{Excess noise induced on CV-QKD in a DWDM network}

\label{sec:noisesources}

We have seen in the previous section how excess noise $\xi$ as well as the transmission $T$ of the quantum channel could be estimated in the context of CV-QKD.

\begin{figure}[htb!]
\centering
 \includegraphics[width=0.8\textwidth]{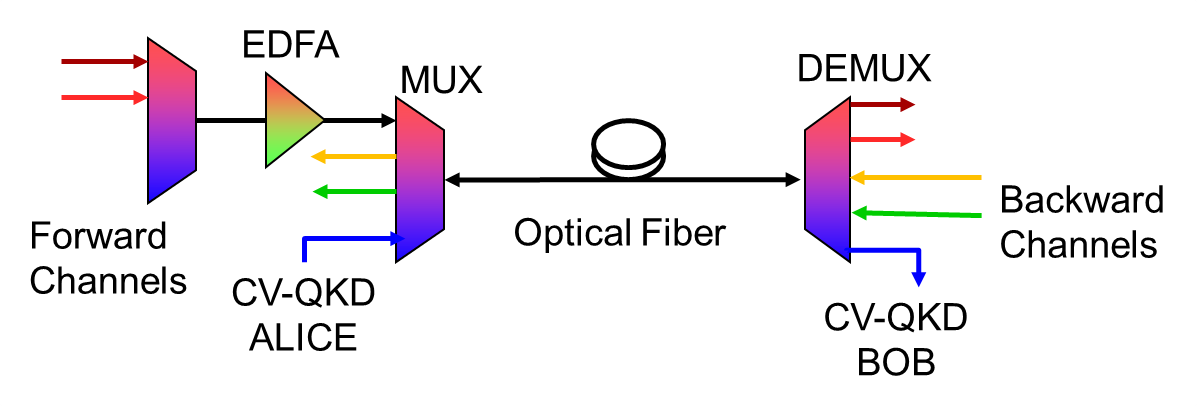}
  \caption{\textbf{Model of the DWDM integration of CV-QKD} To evaluate the different noise contributions arising in a DWDM context, we consider a point-to-point 25 km CV-QKD link multiplexed with  both forward and backward classical channels. (EDFA Erbium Doped Fiber Amplifier,  MUX- multiplexer, DEMUX- Demultiplexer.) }
   \label{figure:DWDMsetup}
    \vspace{-4mm}
\end{figure}

In this section we would like to make an ab-initio analysis of the impact of intense DWDM classical channels on the excess noise and thus on the performance of a CV-QKD system.
To perform this analysis, we consider a relatively generic optical network environment in which the CV-QKD system could be deployed: the CV-QKD system is multiplexed with forward (from Alice to Bob) and backward (reverse) propagating DWDM channels,  by using MUX and DEMUX passive components. In addition, an erbium amplifier is used to regenerated the classical channels in the forward direction, cf Fig.\ref{figure:DWDMsetup}

\paragraph{Assumptions\\}
We want to assess the amount of noise associated to several different process that can contribute to the excess noise, in order to identify the main source(s) of noise. For this we will make the following assumptions:

\begin{itemize}

\item We consider multiplexed quantum and classical optical communication through an optical fiber of 25 km of length.

\item Attenuation coefficient, $\alpha$, of the optical fiber is taken as 0.2dB/km.

\item The clock rate of the CV-QKD system is 1 MHz, and the temporal length of the homodyne detection mode (duration of local oscillator pulses) is of $50$ns.

\item The wavelengths of the quantum channel and of the DWDM classical channels are distinct, located in the C band.

\item The mean number of photon in the local oscillator mode is $\langle n_{LO} \rangle = 10^8$.

\item The launch power of classical channels is of 0dBm (1mW).

\item DWDM multiplexer and
de-multiplexer (MUX and DEMUX) have a 100GHz (0.8nm) separation
between channels.

\item MUX and DEMUX are assumed to have -40dB of isolation between two adjacent channels,  -80dB isolation between non-adjacent channels and negligible
insertion loss.

\item As a simplification, $\eta_B$ the transmission efficiency of Bob, is taken as unity in the  evaluations (this constitutes a worst case assumptions for the estimation of excess noise at Bob).

\end{itemize}

%\paragraph{List of potential noise sources\\}

Based on the above practical assumptions, we have studied quantitatively the amount of excess noise  that could be due to different noise processes.  We will moreover compare these different potential sources of excess noise to the ``system excess noise'' of a CV-QKD system (i.e. systematic excess noise that is due to the CV-QKD system itself, for example because of a limited common mode rejection ratio of the homodyne detection). 
System excess noise is typically larger that $1 \times 10^{-3}$ $N_0$ at Bob, which corresponds to one of the smallest experimental value reported so far \cite{jkl:optexp}. As a consequence, we will consider that noise processes leading to excess noise significantly below $1 \times 10^{-3}$ $N_0$ can be neglected. The analysis of noise contributions that can be neglected in our experiment is given in appendix A.

%\textcolor{red}{System excess noise of $1 \times 10^{-3}$ $N_0$ at Bob is the smallest experimental  value reported so far \cite{jkl:optexp}. Any noise process leading to excess % noise below $1 \times 10^{-3}$ $N_0$ is neglected. Details of such noise contributions are given in Appendix A.} 

%Comparing with the system excess noise of  $1 \times 10^{-3}$ $N_0$ at Bob, which is the smallest experimentally determined value reported so far \cite{jkl:optexp}, any noise process leading to excess noise significantly bellow $1 \times 10^{-3}$ $N_0$ is neglected. The detail of  noise contributions that are typically negligible is given in Appendix A.  

As analyzed in \cite{qi:njp10}, noise photons impinging on the
signal port of the homodyne detection can either be matched or
unmatched with the local oscillator spatio-temporal mod. As the
local oscillator contains $10^8$ photons per mode, this implies
that the photocurrent associated to one matched photon is 80dB
larger than for one unmatched photon, which illustrates the
``built-in'' filtering property associated to coherent detection.  One  point  to mention here is that the amount of noise induced by matched and unmatched photons on homodyne detector is independent of the  quantum signal whether it is  realized with coherent or squeezed or entangled states.
%Since the impact of unmatched photons is strongly reduced by this homodyne detection ``coherent filtering'', the associated excess noise can be neglected and we can focus on excess noise from  % photons that are mode matched with the homodyne detection mode.

We can moreover relate, under the realistic assumption that the statistics of matched noise is chaotic \cite{qi:njp10}, the excess noise (at Bob side, i.e. at the output) $\xi_{out}$ to the mean number of matched noise photons $\langle n_{noise} \rangle$ impinging on Bob device by:
\begin{equation}
\xi_{out}= 2 \eta_B \, \langle n_{noise}  \rangle
\label{photontonoise}
\end{equation}

%After detailed analysis ( given in appendix A) of  contribution to excess noise by different noise sources, it can be seen that Spontanious Raman Scattering (SRS) is the main source of noise in DWDM environment. It is an inelastic scattering process during which scattered photons get converted into photons of either longer  or shorter wavelength, respectively called Stokes and Anti-Stokes scattering. As already notice in \cite{peters:njp09}, SRS is the dominant source of noise for QKD in a DWDM environment, as long as the fiber length is beyond a few km (which is the case in our experimental setup). We will analyze the characteristics and impact of SRS on CV-QKD in a DWDM environment in the next section.  

As already notice in \cite{peters:njp09}, Spontaneous Raman Scattering  (SRS) is the dominant source of noise for QKD in a DWDM environment, as long as the fiber length is beyond a few km (which is the case in our experimental setup). It is an inelastic scattering process during which scattered photons get converted into photons of either longer  or shorter wavelength, respectively called Stokes and Anti-Stokes scattering.  Anti-Stokes scattering is less probable than Stokes. Therefore, in order to minimize the amount of noise due to Raman scattering, it is preferable to place the quantum channel at a wavelength lower than the ones of the classical channels.  We will assume here that this design rule has been followed so that we only need to focus on the effect of Spontaneous Anti-Stokes Raman
Scattering (SASRS) photons on CV-QKD system. In the following, we will analyze the impact of  SASRS on CV-QKD.  

The mean number $ \langle N_{SASRS} \rangle $ of noise photons generated by SASRS per
spatio-temporal  and polarization mode, and matched with the homodyne detection mode, at the input of Bob's QKD system can be expressed as~\cite{eraerds:njp12, qi:njp10}:

\begin{equation} \langle N_{SASRS} \rangle =\frac{1}{2}  \left[
\frac{\lambda^3}{h c^2} \beta \eta_{D} \left( P^{in}_{fwd} \, L
e^{-\alpha L}+ P^{in}_{bwd} \, \frac{1-e^{-2\alpha
L}}{2\alpha}\right) \right]
 \label{raman}
\end{equation}

Where, $P^{in}_{fwd/bwd}$ is the input power of the classical
channel sent either in the  forward  or  backward direction,
$\lambda$ is the wavelength of the quantum channel, $h$ is Plank's
constant, $c$ is the speed of light, $\beta$ is the Raman
scattering coefficient, $\eta_{D}$ is transmittance of DEMUX
(Add-Drop Module) place at Bob side, $L$ is the channel distance
in km and $\alpha$ is fiber attenuation coefficient. In equations
Eq.\ref{raman}, the factor $1/2$  is due to the polarization mode
selectivity of local oscillator.

We can then use Eq.\ref{photontonoise} to evaluate the excess
noise at the output and rescale it by $  \eta_B \eta_D  T$ to
obtain the expression for $\xi_{in}$, the excess noise at the
input, induced by Spontaneous Anti-Stokes Raman Scattering.
\begin{equation}
\xi_{in, SASRS}= 2 \, \langle N_{SASRS} \rangle/ \left( \eta_D  T\right)
\label{xiinsrs}
\end{equation}

From this expression, we can evaluate that a 0dBm forward propagating channel would induce an excess noise of $\sim
1.3\times10^{-3}$ $N_0$ over a 25 km CV-QKD link (assuming $\beta=3 \times 10^{-9}$/km.nm and $\lambda =1531.12nm$) while the excess noise induced by a 0dBm backward propagating channel
 is found to be $\sim 1.6\times10^{-3}$ $N_0$. Comparing this excess noise to other noise sources listed in Appendix A, it can be seen that Raman scattering is the dominant source of noise.
 In the next section, we will experimentally characterize  SASRS noise.

\section{Experimental characterization of Spontaneous Raman Scattering in DWDM environment}

 In order to validate the prominence of Raman Scattering as source of noise, we have conducted experiments to estimate the value of the Raman scattering coefficient,
$\beta$, and to validate experimentally the validity of
Eq.\ref{xiinsrs} in order to predict the amount of excess noise
induced by Spontaneous Raman Scattering on  homodyne detector.

\begin{figure}[htb!]
                \begin{subfigure}[b]{0.4\textwidth}
                \includegraphics[width=\textwidth]{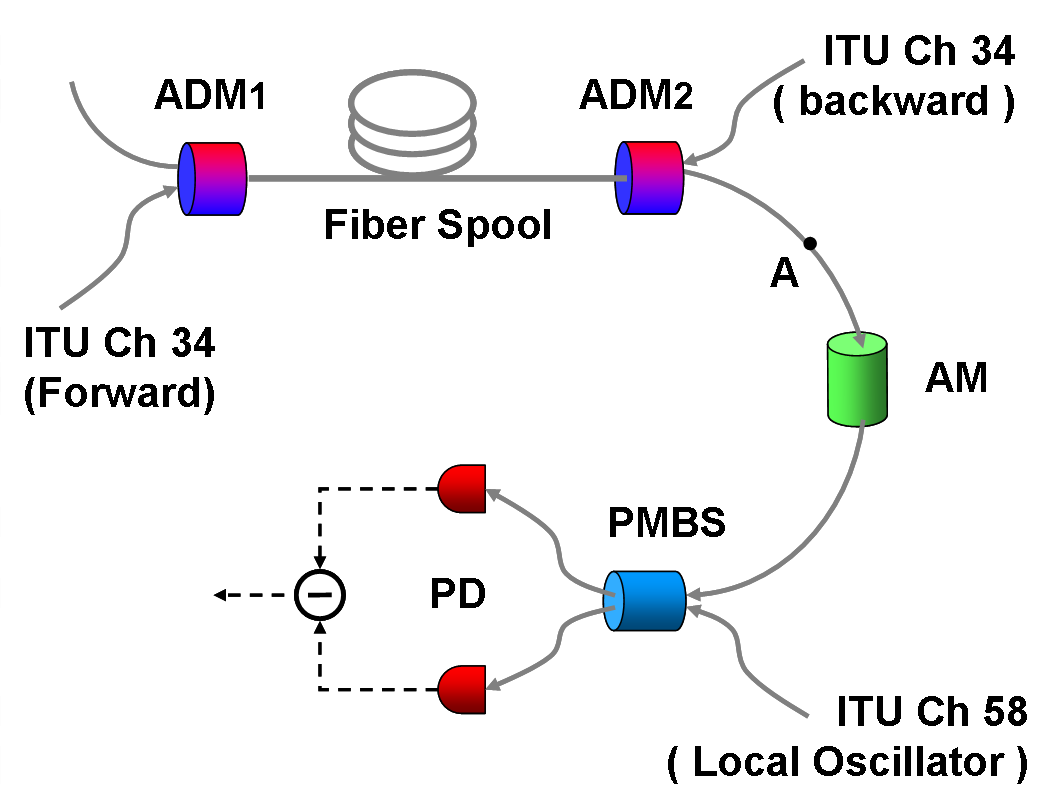}
                \caption{}
                \label{figure:RamanSetup}
        \end{subfigure}
                \begin{subfigure}[b]{0.6\textwidth}
                \includegraphics[width=\textwidth]{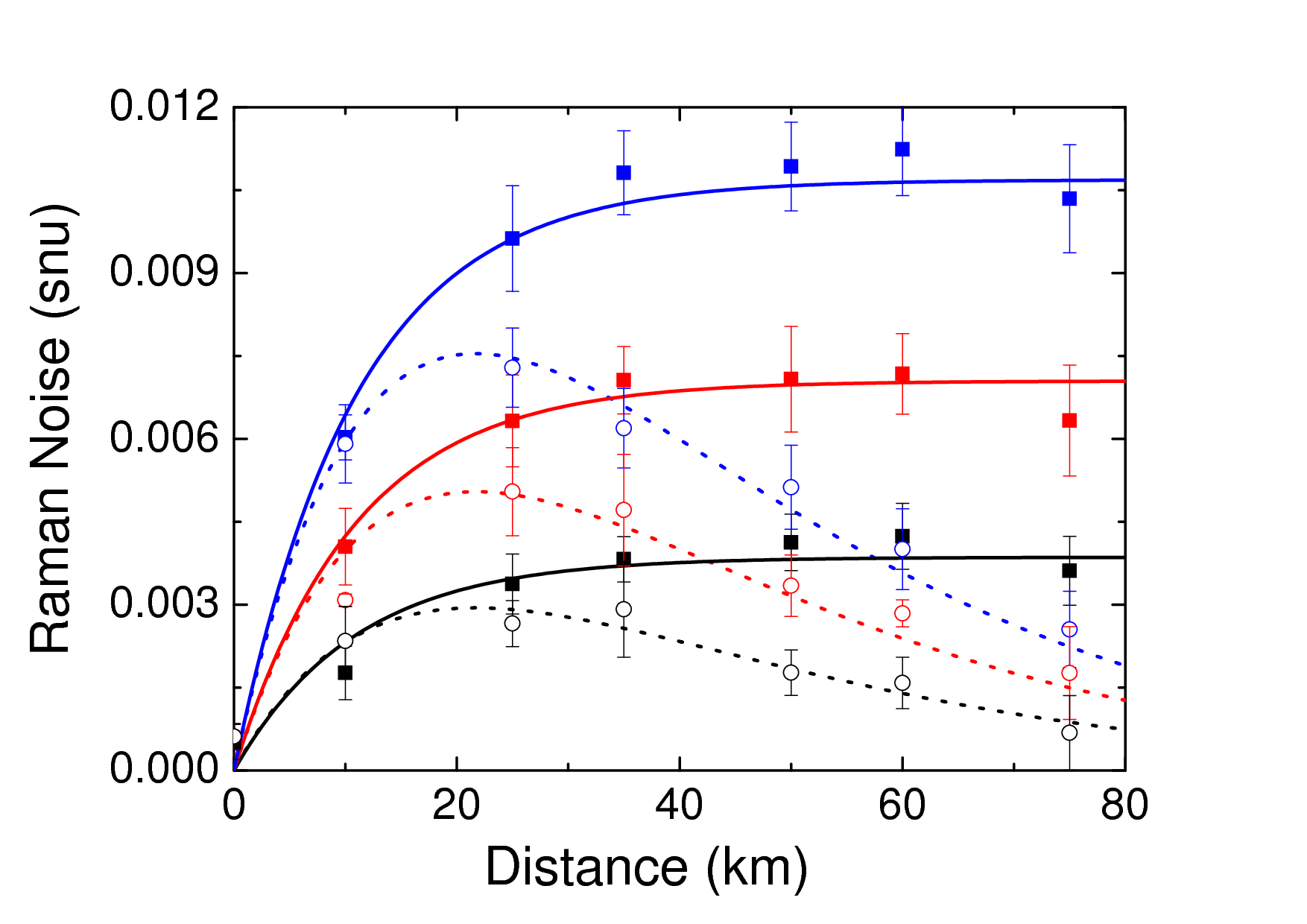}
                 \caption{}
                \label{figure:RamanFWdBwdCalib}
        \end{subfigure}
         \caption{\textbf{Characterization of Raman noise.}\textbf{(a)}:
         Experimental setup. ADM- Add/Drop
         module, AM- Amplitude modulator, PMBS- Polarization
         Maintaining 50/50 Beam Splitter, PD- Photo diodes. \textbf{(b)}: Noise (in excess of shot noise) induced by a classical
          channel of various power on a homodyne  detection. Circles and squares  are experimental data obtained for the classical channel in
  forward and backward directions, respectively. Blue, red and black colors indicate classical
  channel power of 8mW, 5mW and 3mW, respectively. Experimental
  data is fitted  using Eq.\ref{raman} in forward (dotted) and backward (solid)
  directions.}
 \label{figure:RamanClib}
\end{figure}

 The experimental setup is shown in
Fig.\ref{figure:RamanSetup}. Classical channel at 1550.12nm (ITU
channel 34) is multiplexed either in forward of backward direction
into the fibre through  Add and Drop Modules (ADM1/ADM2)  and we
perform noise measurement with a homodyne detector. Add and Drop
Modules are DWDM elements that allow to multiplex/demultiplex
(add/drop) a particular wavelength to/from an optical fiber
channel on which it is placed. ADMs exhibits comparatively less
insertion loss ($\approx$ 0.5dB) than WDM modules ($\approx$2dB).
Moreover, we have obtained cross channels isolation of -46dB
between adjacent channels and  -96dB between non-adjacent
channels. The wavelength of the classical channel is scanned over
the entire C band (1530nm - 1565nm) with 5mW input power.  Raman
scattered photons at the wavelength 1531.12nm (ITU channel 58) of
the quantum channel are collected through the Add/Drop port of
ADM2.

We have characterized the Raman scattering coefficient, $\beta$,
by measuring the intensity of backscattered photons, from a fiber
spool of 25 km, using a power meter (model NOVA II, OPHIR
optronics) at the point A in Fig.~\ref{figure:RamanSetup} and an
ADM of bandwidth 0.8nm. We can use Eq.\ref{raman} to relate the
power of Raman backscattering to the value of the coefficient
$\beta$, whose measurement (that depends on the wavelength of the
classical channel) varies from $ 1.5 \times 10^{-9}$/km.nm. to $
3.1 \times 10^{-9}$/km.nm  and agrees with that given
in~\cite{eraerds:njp12}.

%\begin{figure}[htb!]
%\centering
% \includegraphics[width=0.8\textwidth]{RamanTiming.png}
%  \caption{\textbf{Characterization of Raman noise: timing diagram.}
%  }
%   \label{figure:DWDMsetup}
%    \vspace{-4mm}
%\end{figure}

For the characterization of the excess noise induced by Raman
scattered photons on the homodyne measurement we have used a specific technique to evaluate both total noise and shot noise variance.
 An amplitude modulator (AM) is used to close the signal port of homodyne detector
during the shot noise measurement and is kept  open for total
noise measurement. In order to minimize the effect of homodyne
output drift, shot noise and total noise measurements are taken in
every alternative intervals. Forward and backward Raman noise
variance are measured for different fiber channel lengths with
various classical channel launch power. Results are as shown in
Fig.\ref{figure:RamanFWdBwdCalib}. As we can see, in the forward
scattering direction Raman noise reaches a maximum at $1/\alpha
\sim$ 21km, where $\alpha$ = 0.046 is the fiber attenuation  per
km, and then decreases along with classical channel power. In the
backward direction, noise reaches a saturation level as the
distance increases. The experimental data is fitted using
respective parts of Eq.\ref{raman} and found in good agreement
with the theory.

\section{Excess noise on CV-QKD operated in DWDM coexistence regime: experimental set-up}

To experimentally measure  the excess noise induced by multiplexed
DWDM channels, we have inserted a CV-QKD system in a DWDM
test-bed, and have used a dedicated  scheme for excess noise
acquisition, minimizing system noise associated to temporal
drifts, so that DWDM-induced noise could be resolved with enough
precision. We start by a description of our CV-QKD set-up and then
detail our acquisition scheme.

\paragraph{CV-QKD experimental set-up\\}
 Our CV-QKD system implements the Gaussian
Modulated Coherent Protocol GG02\cite{Grosshans:prl02} and uses a
externally modulated DFB laser at 1531.12nm to general pulses of
temporal width 50ns at a repetition rate of 1MHz. These pulses are
split on a  90/10 beam splitter into local oscillator and signal
pulses. Signal pulses are strongly attenuated (to the level of a
few photons per pulse) and their quadratures are Gaussian
modulated using amplitude and phase modulators, with quadrature
variance $V_A$. Local oscillator and signal  are time multiplexed
(200ns delay) and polarization multiplexed, before being sent to
Bob through the fiber channel. At reception, on Bob side, signal
and local oscillator pulses are polarization and time
de-multiplexed. Detailed description of the setup is given in
\cite{jkl:optexp}. The quadrature information is retrieved by
using a balanced homodyne detector of electronic noise -25dB below
the shot noise.  The intensity of the local oscillator  is set in
order to  have a mean number of $10^8$ photons per pulse at Bob.
The input voltage range of the data acquisition card is set
sufficiently low ($\pm$1 Volts) to obtain a good resolution for
the homodyne output measurement, which reduces the electronic
noise down to 0.3$\%$ of shot noise. Such setting however could
open a door for recently proposed saturation attack on CV-QKD
system \cite{saturation:HAO}, but we will not be considering this
issue, or other issue related to side-channel attacks here.

\begin{figure}[htb!]
\centering
 \includegraphics[width=0.9\textwidth]{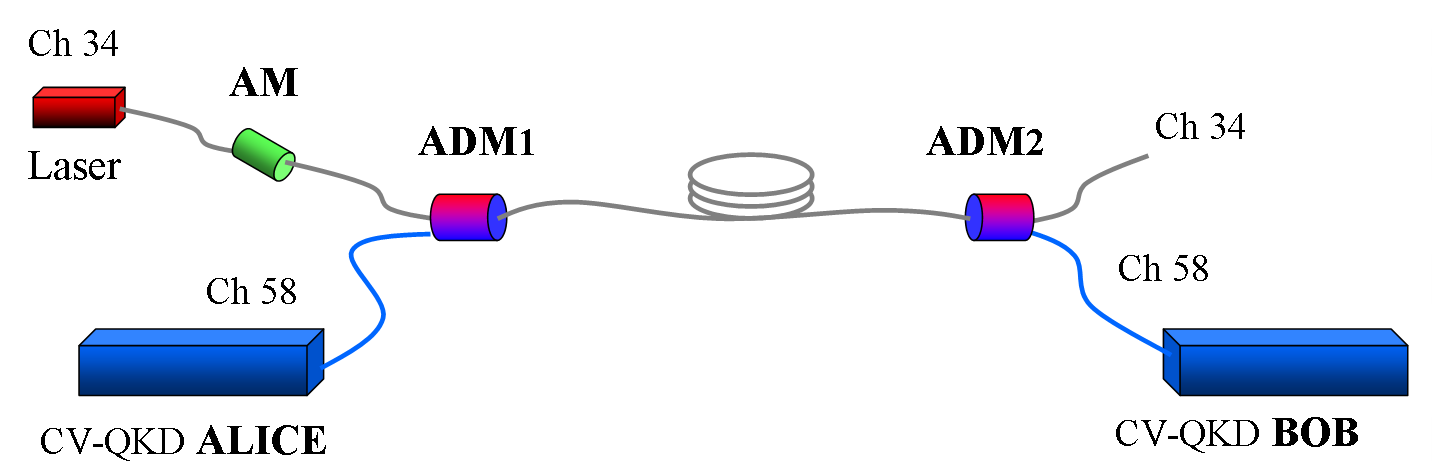}
 \vspace{-2mm}
  \caption{{\bf Setup for excess noise measurement for CV-QKD system operated in coexistence
with one DWDM intense channel}. Add-Drop Modules (ADM 1 and 2) are
used to add and drop, respectively,  the quantum channel to and
from the optical fiber. Amplitude modulator (AM) is used to switch
off the classical channel, while Alice output signal is
synchronously blocked by the AM inside Alice CV-QKQ system. The
figure represents the set-up with a forward propagating classical
channel.  When the classical channel is operated in backward
configuration, the output of the AM is connected to the input of
ADM2 (instead of the input of ADM1). \vspace{2mm} }
\label{figure:setup}
\end{figure}

To perform shot noise measurement (Eq.4) Alice blocks the signal
pulses at emission with her amplitude modulator while a second
amplitude modulator, placed on the classical channel (in green on
Fig.\ref{figure:setup}) is used to block the optical input of the
multiplexed classical channel. On the other hand, when both
quantum and classical signals are multiplexed on the same fiber,
we say that the ``total noise'' variance (Eq.4) is being measured.
In order to limit the impact of statistical fluctuation in
variance estimation \cite{jkl:natphot13}, windows of size $10^8$
pulses were used to estimate the quadrature measurement variances
both for shot noise and for total noise.

We have integrated the above described CV-QKD setup into a DWDM
environment and the experimental setup is depicted in
Fig.~\ref{figure:setup}. We have used a wavelength tunable
continuous laser (model TLS-AG from Yenista) for the classical
channel. The  wavelength of the quantum channel is set at 1531.12 nm (ITU channel 58) so that the quantum channel would be in the Anti-Stockes configuration with respect to any classical channel in the C band \cite{choi:njp11}. The wavelength of the classical channel is set at
1550.12nm (ITU channel 34) based on the choice of available ADMs in the laboratory. It would be possible to select a quantum channel wavelength close to the classical channel, as illustrated in \cite{eraerds:njp12},  in order to further minimize the Raman induced noise. Both channels are multiplexed and de-multiplexed
to and from the optical fiber spool by means of ADMs.  An
additional bandpass filter (not shown in Fig.~\ref{figure:setup})
had also placed (before the ADM) on the classical channel in order
to remove sidebands (such filtering would naturally be present if
a multi-channel MUX had been used, as shown in
Fig.\ref{figure:DWDMsetup}).

%The wavelength of the classical channel is set at
%1550.12nm (ITU channel 34) and quantum channel is set at 1531.12nm
%(ITU channel 58).  Both channels are multiplexed and de-multiplexed
%to and from the optical fiber spool by means of ADMs.  An
%additional bandpass filter (not shown in Fig.~\ref{figure:setup})
%had also placed (before the ADM) on the classical channel in order
%to remove sidebands (such filtering would naturally be present if
%a multi-channel MUX had been used, as shown in
%Fig.\ref{figure:DWDMsetup}). \textcolor{red}{Our selection of the quantum and classical wavelengths are purely based on the channel specifications of ADM and other passive optical elements  available in the laboratory. It is possible to select quantum channel wavelength close to classical channel, as selected in\cite{eraerds:njp12}, such that Raman induced noise is minimum. }

In our derivation of the secret key rate, we have opted for a
conservative approach and have considered that the losses
associated to the ADMs are part of the fiber channel and not of
the QKD system. This approach is conservative because it does not
require to assume that the ADMs insertion loss are calibrated (and
this calibration trusted). Moreover, this approach allows to
directly take into account possible variations of the insertion
loss of the ADMs, due to wavelength drift of the CV-QKD laser
around the transmission window of the ADMs,  in the estimation of
channel transmittance $T$ (Eq.2).

\begin{figure}[htb!]
        \begin{subfigure}[b]{0.6\textwidth}
                \includegraphics[width=\textwidth]{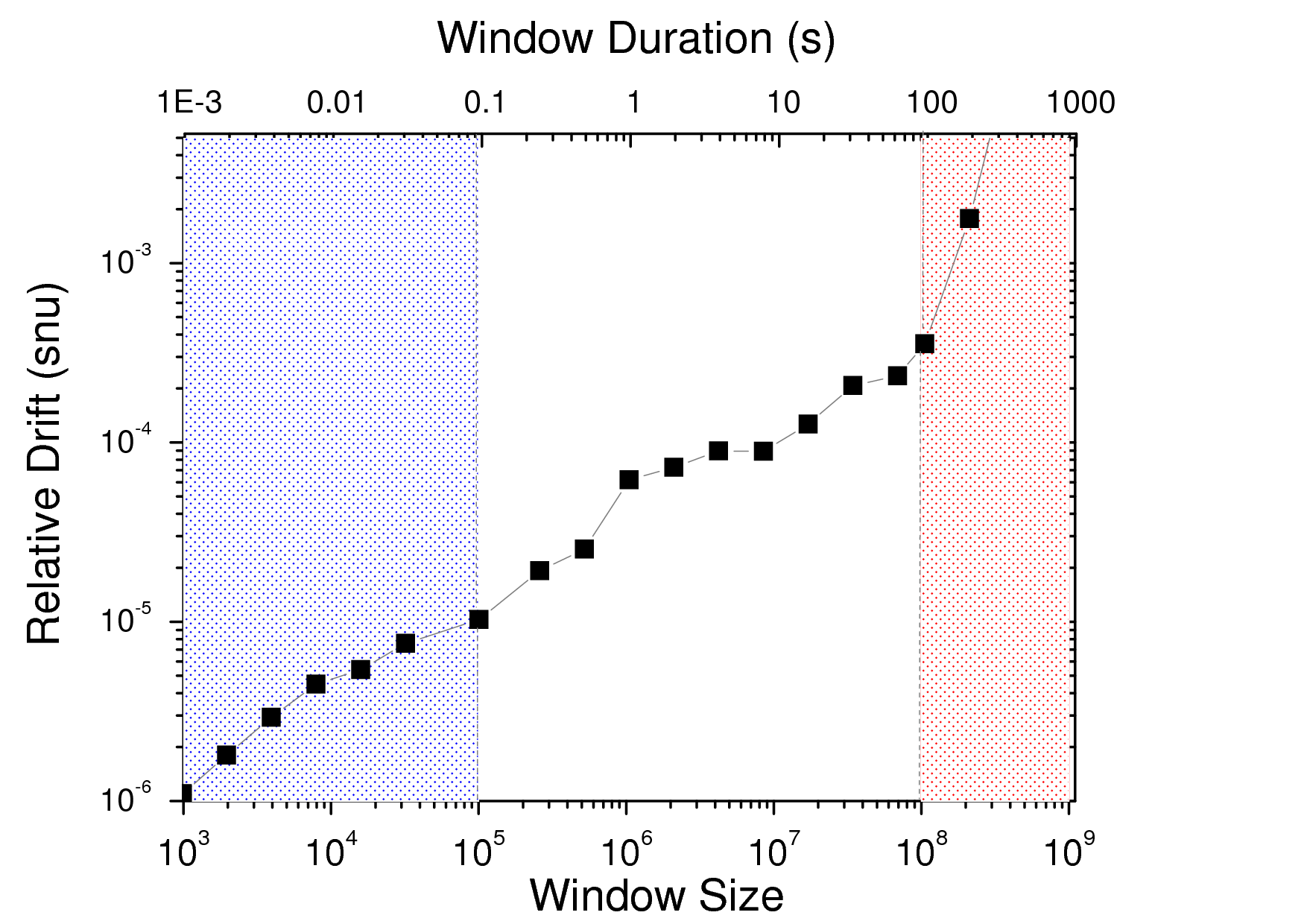}
                 \caption{}
                \label{figure:TemperatureDrift}
        \end{subfigure}
                \begin{subfigure}[b]{0.4\textwidth}
                \includegraphics[width=\textwidth]{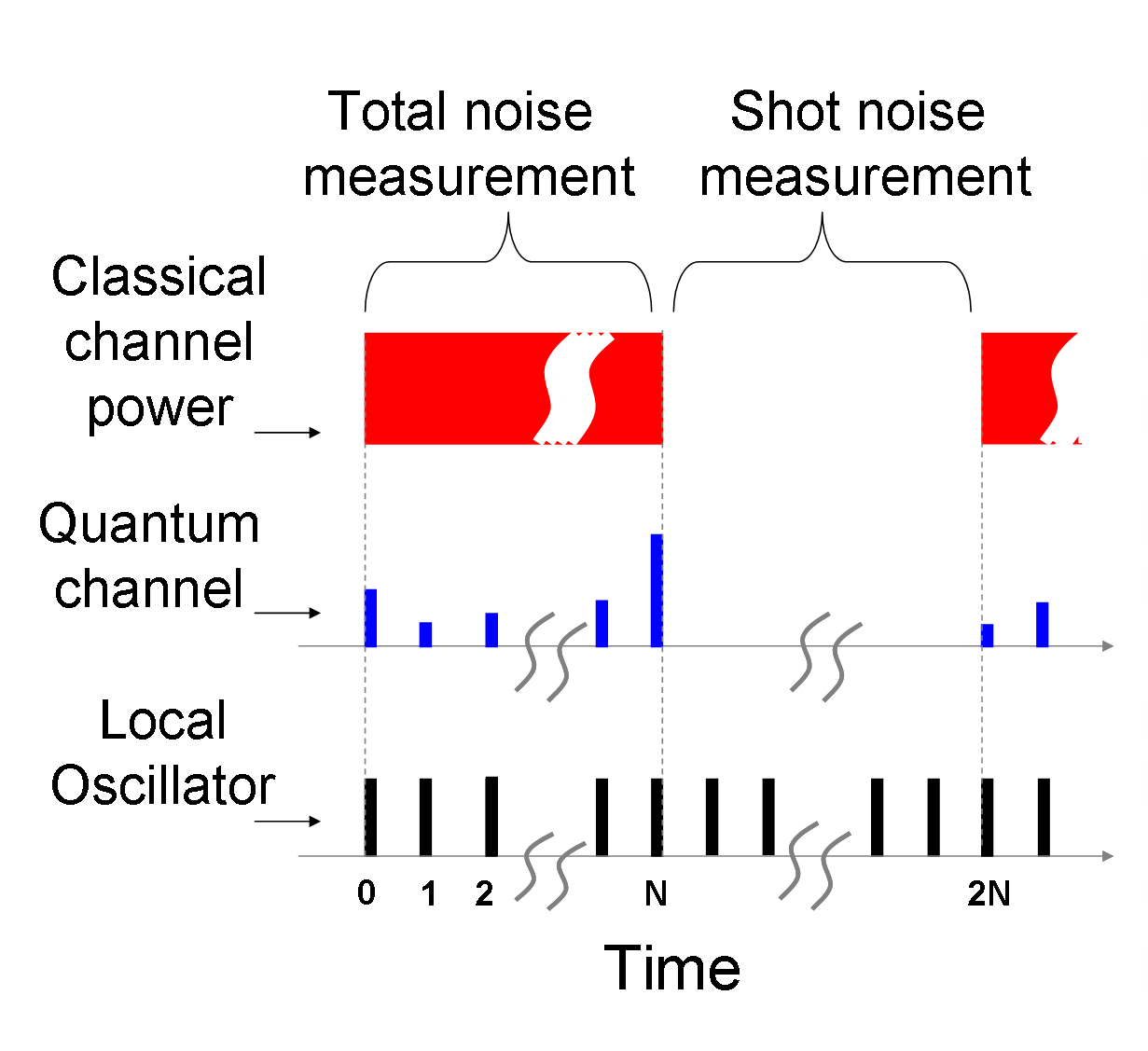}
                \caption{}
                \label{figure:timing}
        \end{subfigure}
        \caption{\textbf{(a)}: \textbf{Drift of homodyne output}.
        Relative drift between shot noise variance measurements in consecutive blocks, as a function of data window size.
       Data window size ranging from 1ms to 200s have been used (corresponding to window size ranging from
        $10^3$ to $2\times10^8$ homodyne measurements). At short time scales (below $10^5$ measurements) the relative drift is negligible (region shaded in blue), while it becomes comparable to the system noise ($10^{-3} N_{0}$ for time scales of several seconds (regions shaded in red).
         \textbf{(b)}: \textbf{Timing diagram}. Data acquisition is divided in relatively short periods (N =$10^5$ pulses, corresponding to 100 ms). Shot noise and total noise measurements are performed alternatively on each consecutive period, in order to limit the relative drift.}
   \label{figure:SetupANDtiming}
\end{figure}

\paragraph{Excess noise measurement and drift compensation\\}

To evaluate the excess noise with a good statistical precision
large data blocks  should be used. This is indeed an important
issue when one wants to deal properly with finite-size issues
\cite{jkl:natphot13}. In our case, we should typically compute
estimators on data blocks of size $10^8$ in order to have a
statistical fluctuations around $10^{-4} N_0$ and thus below
system excess noise. However, the value of homodyne output can
drift with time (essentially due to temperature fluctuations that
modify the balancing conditions of the homodyne detection). This
temporal drift results in an additional noise that depends on data
block size, as one estimates shot noise and total noise on two
consecutive data blocks, as we can see from
Fig.\ref{figure:TemperatureDrift}
 If large window size (200s) were used to measure consecutively total noise and shot noise, then the drift could generate an additional noise of the order of $1.5\times10^{-3} N_0$ and thus strongly affect the precision of our DWDM-induced noise measurements.

One way to mitigate this effect is on the other hand to use
smaller data blocks so that the noise induced by the relative
drift becomes negligible at this timescale. Using this principle,
we have used data blocks of size $10^5$ pulses (100ms)
alternatively for the acquisition shot noise and total signal
measurements. Excess noise estimation on data blocks of $10^{8}$
acquisition is then obtained by concatenating  $10^3$ data blocks
obtained at the 100 ms timescale, reducing the relative
statistical uncertainty.

%One way to mitigate this effect is on the other hand to use
%smaller data blocks to compute estimators, so that the relative
%drift between blocks (and the induced excess noise) becomes
%negligible compared to the actual excess noise  Using this
%principle, we have used two data blocks of $10^5$ pulses (50 ms)
%to alternatively compute estimators for shot noise and total
%signal variances performed the subtraction of these estimators in
%order to compute an estimator for the excess noise. Excess noise
%estimation on data blocks of $2 \time10^8$ acquisition is then
%obtained by averaging $10^3$  results obtained at the 100 ms
%timescale, reducing the relative statistical uncertainty, which is
%an important issue when one wants to deal properly with
%finite-size issue \cite{jkl:natphot13}.

To perform shot noise measurement in our set-up, the emission of
Alice and the emission of the laser source on the classical
channel were shut down simultaneously during 100ms with respective
amplitude modulators, as depicted on Fig.~\ref{figure:timing}. The
relative drift at the 100ms timescale is around $10^{-5} N_0$  (measured at Bob), can
be neglected since it is significantly smaller than the CV-QKD
system excess noise ($10^{-3} N_0$ in our case, at Bob).

As proposed in \cite{switch:Paul}, it is possible to use another scheme in order to perform time-resolved shot noise measurements:
it consists in using an amplitude modulator located at the entrance of Bob to modulate $\eta_B$, and to use Eq.3 (for different value of $\eta_B$) to evaluate both shot noise and excess noise variance. This approach is feasible but would interfere with some routines already implemented in our CV-QKD implementation, related to Alice-Bob data synchronization and phase tracking. We have therefore not opted for this design in our DWDM test-bed but plan to do so it in future field trials.

\section{CV-QKD experimental coexistence tests: results and analysis }

We have operated our experimental test-bed  of CV-QKD multiplexed
with one DWDM classical channel (described in the previous
section), at 25km, 50km and 75km with a classical channel power after the ADM 
varied from 0mW to 8mW. For each experimental run,  transmission
$T$ and excess noise $\xi$ were evaluated from the experimental
data, using the Eq.1-4.

\begin{figure}[htb!]
\centering
 \includegraphics[width=0.7\textwidth]{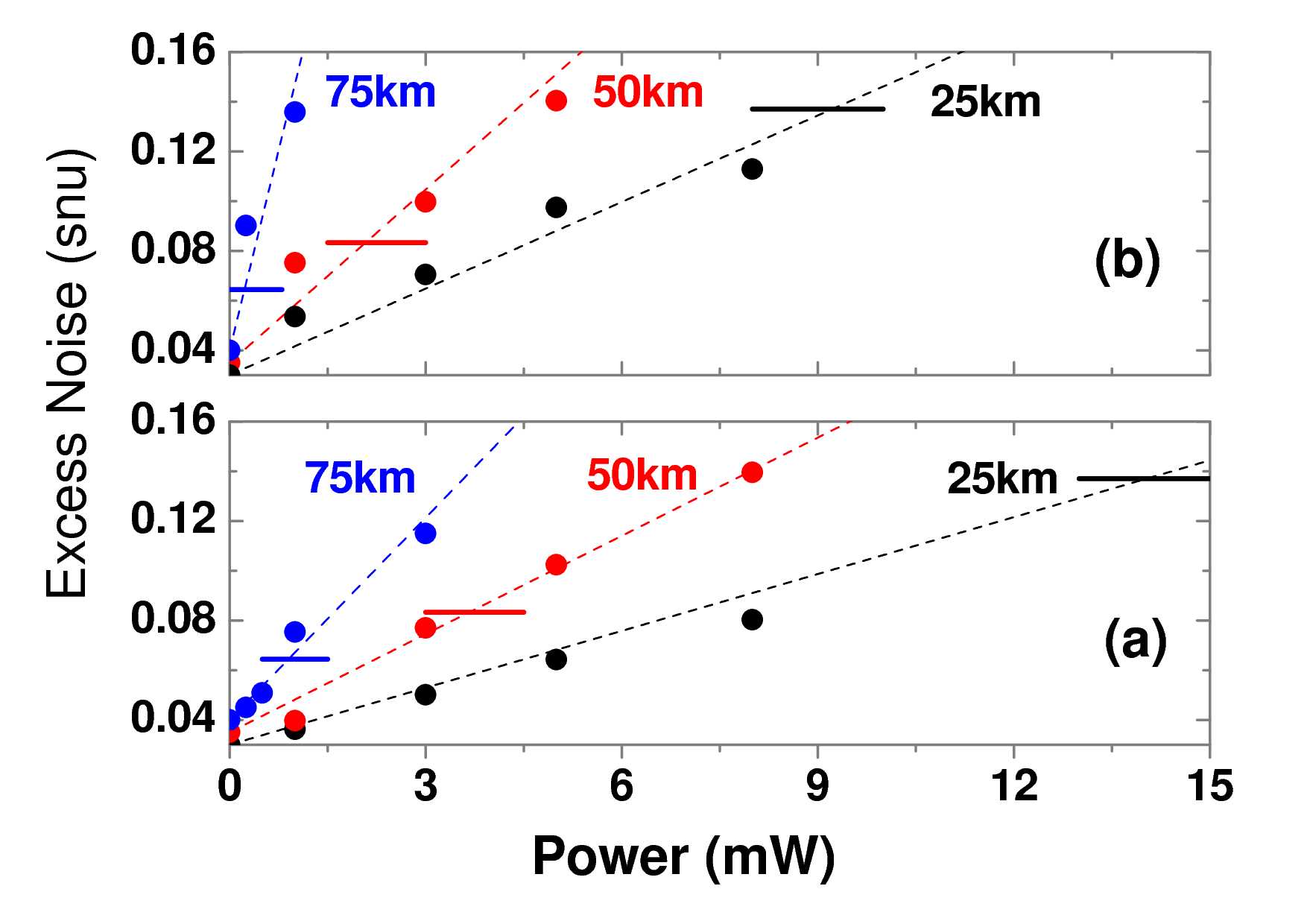}
 \vspace{-2mm}
  \caption{\textbf{Excess noise measurements in forward (a)  and backward  (b) channel configuration}.
  Black, red and blue data points are the  excess noise  evaluated at Alice
  for channel length of 25km, 50km and 75km, for different classical channel power.
    Dashed lines indicate the expected excess noise curve and solid horizontal
  lines are null key threshold for respective channel distance.
  See text for details.
\vspace{2mm} } \label{figure:Result}
\end{figure}

 The measured excess noise  at the output of Alice as a function of classical power are displayed in Fig.\ref{figure:Result}.
 We compare these experimental values to the expected excess noise, i.e. the  sum of the system excess noise
 (that is calibrated to be $0.03 \, N_0$ in our case, at Alice) with the noise associated
 to spontaneous Raman emission, that can be computed from Eq.\ref{raman} and Eq.\ref{xiinsrs}. We in particular expect the excess noise to be a linear function of the launch power.

% \vspace{-5 mm}

We also position the null key thresholds on
Fig.\ref{figure:Result},
 i.e. the maximum excess noise that can be tolerated in order to be able to obtain a positive secret key rate.
 Assuming collective attacks and 0.95 reconciliation
efficiency, the null key threshold  for 25km is 0.137$N_0$,
0.083$N_0$ for 50km and 0.064$N_0$  for 75km.

It can thus be seen that a positive key rate can be obtained for
classical channel power up to 14mW at 25km,  3.7mW at 50km and
0.89mW at 75km in forward configuration whereas  in backward
direction admissible classical power drops to 9.3mW, 2mW and
0.23mW, respectively. The secure key rate (under collective attacks) has been calculated from the evaluated excess noise $\xi$ and transmission $T$  taking into account finite-size effects with our data block size of $10^8$. Worst-case estimators for the excess noise (with 3 sigma of deviation) have be used, following the analysis \cite{jkl:natphot13}. With single 0dBm channel at distance 25km, the key rate is 
 24.11kb/s  in forward and 22.98kb/s in backward direction. In 50km channel length the key rate drops to 3.16kb/s and 2.27kb/s, respectively. We have also obtained a positive key rate of 0.49kb/s at 75km by reducing the classical channel power (while considering classical channel receiver sensitivity below -25dBm) to -3dBm in forward and -9dBm in
backward direction. One important thing to point at here is  the
yield (secure key bit per QKD signal pulse) of CV-QKD system in
WDM environment. In our experiment with 0dBm classical channel
over 25km the yield is of $485\times10^{-4}$ bits/pulse which is
two order of magnitude higher than the recently reported,
$485\times10^{-6}$ bits/pulse, DV-QKD experiment
\cite{Patel:apl14}. On the other hand, the latest DV-QKD system
can be operated at GHz-clock rate, which has still not yet been
demonstrated with CV-QKD systems, currently operated at MHz clock
rate, even though no fundamental barrier prevents to upgrade it to
100 MHz if not GHz clock rate.

\begin{figure}[htb!]
\centering
 \includegraphics[width=0.8\textwidth]{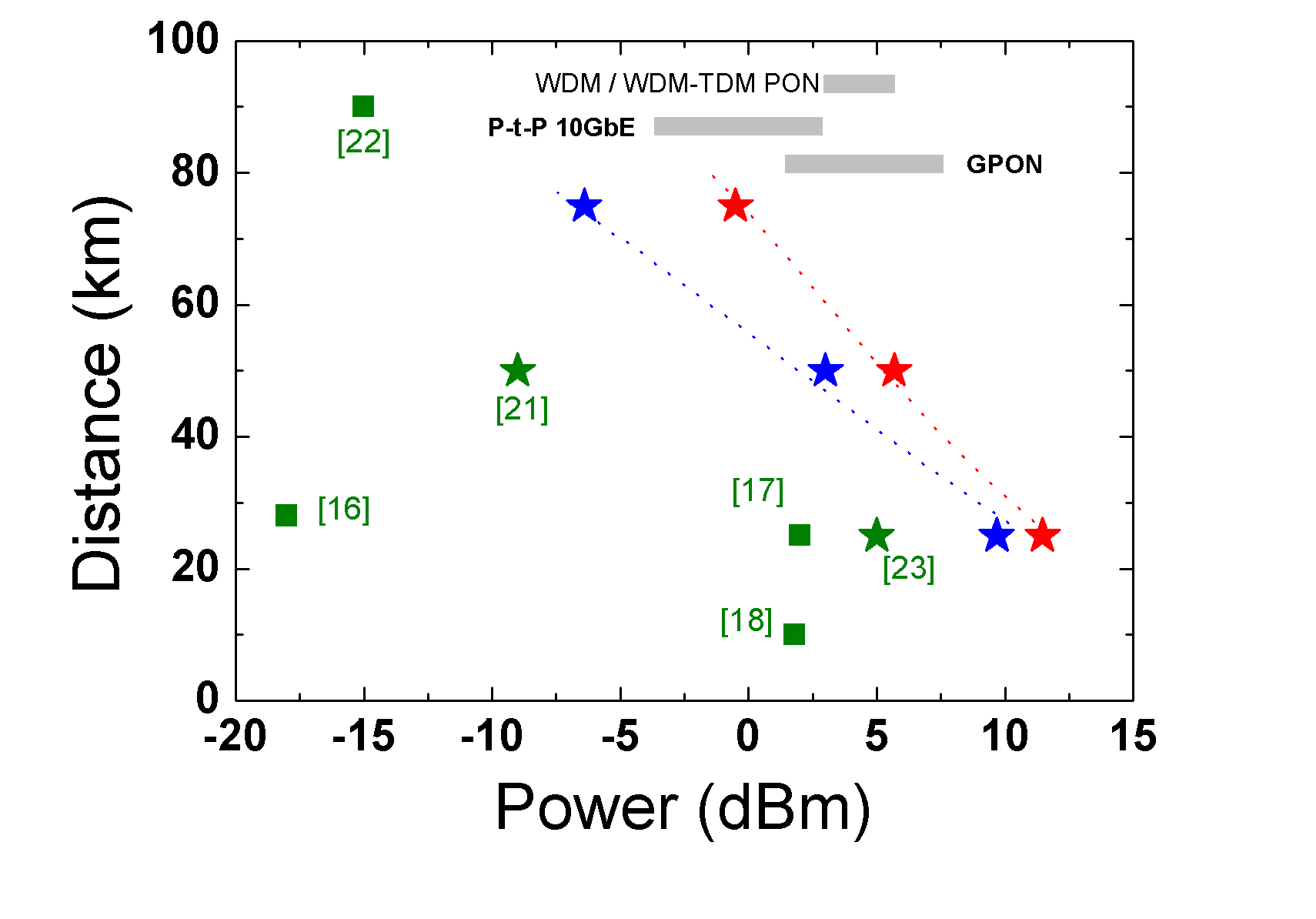}
 \vspace{-2mm}
  \caption{ \textbf{Tolerable classical channel power vs Reachable distance}: Performance of QKD in the context
  of coexistence with classical optical channels.
  Red and blue colors represents our results with a CV-QKD system, in forward and backward classical channel configuration,
  while previous
  works with DV-QKD systems are in Green.
  Stars: experiments conducted in the C-band (DWDM). Squares: experiments conducted in CWDM. The doted red and blue lines are the forward and backward simulation curve for the null key rate in
  the current experiment. Gray bands show transmitter input power range in different standardized optical networks.
\vspace{2mm} } \label{figure:CompareWDMworks}
\end{figure}

\paragraph{Comparison with DV-QKD\\}

To illustrate the strong DWDM coexistence capacity of CV-QKD, we
have made a comparative study with previously  reported DV-QKD
experiments~\cite{Townsend:elelett97, eraerds:njp12, patel:prx12,
choi:njp11, chapuran:njp09, Patel:apl14}, and displayed in
Fig.\ref{figure:CompareWDMworks} a comparison of the reachable
distance of QKD, as a function of the classical multiplexed power
(in CWDM or DWDM, see caption).
In Fig.6, the data points for CV-QKD indicate the maximum reachable distance (null key threshold).  The key rates corresponding to experimental points taken with our CV-QKD system and displayed on Fig. 6 are: 12b/s for 25km; 8b/s for 50km and 9b/s for 75km. Note that  DV-QKD performances mentioned in Fig.6 have also been acquired very close to the null key threshold.  One important thing to note that the different results mentioned in this comparison do not all rely on a unified security analysis. Key rates are derived for security proofs valid against collective attacks in \cite{ choi:njp11,  Patel:apl14, eraerds:njp12}, individual attacks in \cite{Townsend:elelett97} and general attacks in\cite{patel:prx12}, while among these references, only \cite{Patel:apl14} takes finite-size effects into account. As previously explained, we have considered collective attacks and have taken finite-key effects into account for the CV-QKD secure key derivations associated to our experiments.
 
 It can be seen that CV-QKD can
reach longer transmission distances for a given classical channel
launch power. Conversely, for a given transmission distance,
CV-QKD can tolerate noise from multiple classical channels with
typical transmission power of 0dBm. This is particularly true for
25km and 50km transmission distance as shown in
Fig.\ref{figure:CompareWDMworks}. CV-QKD can also be deployed in
coexistence with classical channels of unprecedent power levels-
thanks to the mode selection property of its coherent detection.
This gives CV-QKD an advantage for the integration into different
optical network architectures and in particular into access
networks. Such integration requires, in general, capacity for QKD
to co-exist with classical channels of several dBm of power.  As
it can be seen in Fig.~\ref{figure:CompareWDMworks}, strong
co-existence of CV-QKD would allow its integration  into different
standard passive optical networks such as, for example, Gigabit
PON, 10G-PON and WDM/TDM PON\cite{Netwrok:Power}.
\begin{figure}[htb!]
\centering
 \includegraphics[width=0.8\textwidth]{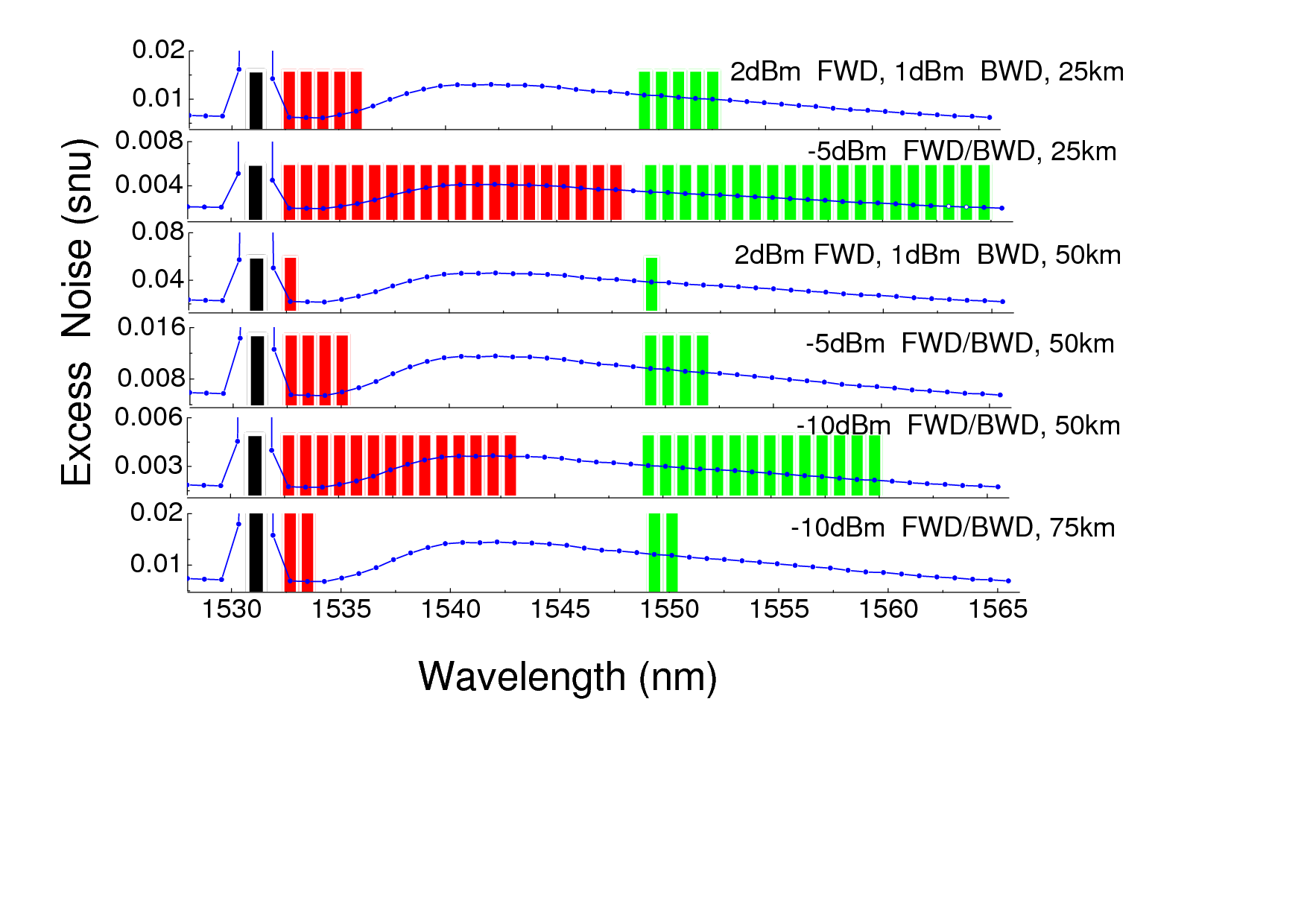}
 \vspace{-2mm}
  \caption{\textbf{Optimized classical channel allocation for CV-QKD in WDM-PON
   network.} In Black: the 1531.12 nm channel allocated for the quantum channel.  Red
   and Green bars represents the backward and forward
   classical channels, positioned on the DWDM wavelength grid . Each blue dot (connected by the blue line) represents the simulated value of the Raman-induced excess noise arising from one backward (of specified power) classical channel onto the quantum channel. Simulated data for excess noise from
   forward channels is not shown.
\vspace{2mm} } \label{figure:OptimAllocation}
\end{figure}

\paragraph{Optimization of Classical channel allocation\\}

In the light of the experimental results and the promising
perspective defined for CV-QKD integration in optical
networks, we have simulated how we could effectively integrate
CV-QKD in some WDM Passive Optical Network architectures
(WDM-PON).
To this effect, we have applied a simple optimization routine to
the integration of CV-QKD system into WDM-PON, that allowed us to
propose classical channel allocations by minimizing the excess
noise induced on CV-QKD.

For a typical access network distance of 25km, we have considered
classical channel allocation in the C band and found that CV-QKD
could coexist with 5 pairs of classical channels (with nominal
WDM-PON channel launch power: 2dBm in forward and 1dBm backward).
Optimization(at a give transmission distance) is performed by
choosing  sequentially the position of the classical channel that
maximize the additional excess noise on QKD, up to the maximum
number of channels compatible with a positive secret key rate.

  If the detector sensitivity on the classical channels
allows,  it might even be realistic to reduce the classical
channel power below the nominal specifications of a WDM-PON
network, while still being able to operate the classical channels.
We have studied the impact of this hypothesis  in
Fig.\ref{figure:OptimAllocation}. We can see for example that 14 pairs of
channels (each with  -10dBm launch power)  could be multiplexed
with one CV-QKD channel at 50km and while 2 pairs of channels
(also with  -10dBm launch power) could coexist with CV-QKD at
75km. These simulation results
 clearly indicate that the strong coexistence capacity of CV-QKD with WDM multiplexed classical channels is likely to play an important role in the integration of QKD into optical networks.

\section {Conclusion}

The success of emerging optical network technologies relies for a
large part on their ability to be seamlessly integrated into
existing infrastructures.  We have demonstrated the successful
co-existence capability of CV-QKD intense (around 0 dBm) classical
channels, in a DWDM configuration.  We have characterized and
studied the influence of the main source of noise: Raman
scattering and have demonstrated experimentally that CV-QKD can
coexist with a DWDM channel intensity as high as 11.5dBm, while
positive key rate could also be obtained with a -3dBm forward DWDM
multiplexed classical channel at 75km.

It can be also  seen that CV-QKD, benefitting from a built-in single mode filtering (associated with the coherent detection) is less affected by DWDM-induced noise photons that the DV-QKD systems tested so far in this regime, and can therefore reach longer transmission distances for a given DWDM classical channel launch power.

These experimental results indicate that CV-QKD, and more generally coherent communications operated at the shot noise limit are a promising technology in order to jointly operated quantum  and classical communications on the same optical fiber network, and can therefore play an important role for the development of quantum communications over existing optical networks.

\section {Acknowledgement}
We thank G\'erard Mouret for his very helpful support for the design of the electronics of the experiment. We also thank Paul Jouguet and S\'ebastien Kunz-Jacques for useful discussions on CV-QKD noise evaluation  and  Zhiliang Yuan for suggesting corrections in Fig.6. This research was supported by the ANR, through the project QUANTUM-WDM (ANR-12-EMMA-0034), by the DIRECCTE Ile-de-France through the QVPN (FEDER-41402) project and by the European Union through the Q-CERT (FP7-PEOPLE-2009- IAPP) project.

\section {Appendix A. Negligible source of noise in our DWDM experiment.}

The excess noise from different sources is calculated based on the assumptions given in section\ref{sec:noisesources}. While spontaneous Raman scattering constitutes the main source of noise, we assess here the impact of other potential sources of noise and show that they can be neglected in our experiment, being significantly below the limit of the smallest observable system excess noise of $10^{-3}$ $N_0$.

\begin{itemize}

\item Imperfect isolation of DEMUX: due to limited isolation between the channels of
DEMUX, light from forward classical channels can leak into the
quantum channel and induce excess noise on the QKD system.
Sufficient cross channel isolation usually permits to eliminate
this noise. With the numerical assumptions listed above (in
particular 0 dBm launch power of classical channels and realistic isolation for
MUX and DEMUX), we can evaluate that the excess noise due to
unmatched leakage photons from non-adjacent channel would lead to
an excess noise of  $\sim6\times10^{-9}$ $N_0$. Channel isolation is typically smaller for adjacent channel, but the excess noise would be in this case approximately
of $\sim6\times10^{-5}$ $N_0$. 

 \item Rayleigh scattering (RS): RS is an elastic scattering
 process, that can in particular lead to back-reflection in an optical fiber (Rayleigh backscattering). The wavelength of the scattered light is identical to that of the incident
light. Rayleigh backscattering can be a major concern for the design of two-way DV-QKD systems
deployed over dark fiber\cite{Subacius: apl05}. In the context of
CV-QKD implemented in WDM environment however
%we can first state that the effect of Rayleigh
% scattering will be more important from backward-propagating classical
% channel (as the intensity is launched near Bob ) than from forward channel.
%However, RS is an elastic process and
 the isolation of DEMUX will act as a efficient protection, as illustrated in the previous item. Indeed, since the intensity of backscattered photon will be in all case very low compared to the channel launch power itself,  the noise contribution due to Rayleigh scattering will be negligible.

%  \item Stimulated Brillouin Scattering (SBS): in SBS, an incident
%  photon is converted into a scattered photon of higher wavelength and a
%phonon.  Due to phase matching consideration, the scattered photon
%propagates in the backward direction. At 1550nm, the scattered
%photon is shifted by $\sim$8.8pm, which means it remains in the same DWDM
%channel~\cite{Chraplyvy:jlightTech90} and cannot generate a significant cross-talk into the quantum channel. The impact of SBS on the
%excess noise is thus negligible.

 \item  Stimulated Brillouin Scattering (SBS): in SBS, an incident
  photon is converted into a scattered photon of higher wavelength and a
phonon.  Due to phase matching consideration, the scattered photon
propagates in the backward direction. At 1550nm, the scattered
photon is shifted by $\sim$8.8pm, which means it remains in the same DWDM
channel~\cite{Chraplyvy:jlightTech90} and cannot generate a significant cross-talk into the quantum channel. 

\item  Guided Acoustic Wave Brillouin Scattering (GAWBS) can be another possible  source of noise in high bandwidth QKD system, where photons from local oscillator pulse scatter to the signal pulse, despite the isolation (temporal multiplexing) and contribute to the excess noise. However, the spectral width of GAWBS  is of approximately 600 MHz \cite{Li:josab14} and the associated noise contribution be essentially cut with a homodyne detection operating around 1 MHz. Moreover the isolation associated to the temporal multiplexing between signal and LO used in our system (200 ns) is much larger than what can be done in a 1 GHz system. As a consequence, while the effect of GAWBS can become a problem with high-speed system (1GHz) bandwidth  \cite{Li:josab14}, it is negligible in our experiment operated at 1 MHz repetition rate.

\item  Four Wave Mixing (FWM): when two or more pumps of different frequency co-propagate in the
optical fiber, by the third order non-linear process, involving
the non-linear coefficient $\chi^3$, additional frequencies are
generated. Depending on the wavelength configuration,
FWM-generated photons may fall in channels on the DWDM grid
adjacent to the classical pump channels ~\cite{peters:njp09}.
Considering a separation of 100GHz between two 0dBm classical
channels, the power generated into a third channel (that could be
the quantum channel) is approximately -81dBm. That would lead to
an excess noise of $\sim 6\times10^{-4}$ $N_0$.

\item Amplified Spontaneous Emission (ASE): Erbium Doped Fiber Amplifiers (EDFA) used in optical networks generate a wide-band optical noise due to spontaneous emission. Noise photons are
generated over the entire C band and thus affect all the DWDM
channels, an in particular the quantum channel. Considering an
amplification gain of 100 of the EDFA and following the analysis
of \cite{qi:njp10}, the excess noise due to ASE, filtered by the
MUX at Alice side (whose side function is to act as a bandpass
filter) is found to be approximately of $6\times10^{-7} \,N_0$.

\item Sideband photons:
photons  generated  by the classical channel laser at the
wavelength of the quantum channel are a source of matched noise
photons. However, the power of these sideband photons is typically
-40dB below the main mode for typical semiconductor DFB lasers.
Like for ASE, the wavelength bandpass filtering property of MUX
can moreover reduce this noise to a great extent. For example,
side band photons from a 0dBm channel adjacent to quantum channel
would lead to an amount of excess noise of
$2.4\times10^{-4}$ $N_0$.

\item Cross Phase Modulation (XPM): in Cross Phase Modulation, power fluctuations of the classical
channels affect  the refractive index of the fiber and it induces
phase fluctuations on the quantum channel. In a typical CV-QKD set-up quantum signal
and local oscillator are time-multiplexed and may thus experience different phase modulation due to XPM. Their relative phase drift can be expressed as
 $\phi=  4 \pi n_2 L_{eff} (P^{s}_{in}-P^{lo}_{in})/ \lambda_c A_{eff}$, where, $n_2= 3\times 10^{-23} m^2/mW$
 is the intensity dependent refractive index of the
 fiber, $\lambda_c$  is the wavelength of classical channel,  $A_{eff} = 83  \mu m^2$ is the effective area and
 $L_{eff} = (1- e^{-\alpha L})/\alpha$  the effective length of the fiber.
 $P^{s}_{in}$ (respectively  $P^{lo}_{in}$) is the launch power of the classical channel
  as it temporally overlaps with the signal (respectively with the local oscillator) pulses.
 The phase drift creates an amount of excess
noise of  $ V_A \, Var(\phi) $~\cite{qi:pra07,
Chraplyvy:jlightTech90}. Considering the less favorable case,
where the classical power would overlap temporal maximally with
the quantum signal pulses ($P^{s}_{in} = P_{in}$), and not at all
with the local oscillator ($P^{lo}_{in} =0$), which could
correspond to a situation where the classical channel is operated
exactly at the repletion rate (1MHz in our case) of the quantum
channel, with a ON-Off Keying type of amplitude modulation and a
mean power of 0 dBm. Then the XPM excess noise contribution after
25 km is found to be $\sim 1.3 \times10^{-5}$ $N_0$ and would
saturate to $\sim 2.8 \times10^{-5}$.

%\item Spontaneous Raman Scattering (SRS) : it is an inelastic scattering process
%during which scattered photons get converted into photons of
%either longer  or shorter wavelength, respectively called Stokes
%and Anti-Stokes scattering. As already notice in \cite{peters:njp09}, SRS is the dominant source of noise for QKD in a DWDM environment, as long as the fiber length is beyond a few km (while FWM could dominate at short distances). We will analyze the characteristics and impact of SRS on CV-QKD in a DWDM environment in the next section.

\end{itemize}


\begin{thebibliography}{50}
\bibitem{NoteKeyRate} The notion of ``distance limitation of QKD '' is, as such, ill-defined. If one wants to establish a comparison between the performance of different QKD systems, a large number of parameters have to be jointly considered, starting with the security model and the secure key rate, together with important more ``practical'' parameters that include channel loss, noise environment and the photodetector technology. The numbers we indicate are approximative estimates of the maximum distance at which QKD could be operated with state-of-the art QKD systems, ten years ago \cite{Stucki:njp02} and now \cite{Lucamarini:oe13}, with a key rate of at least 100 bits/s sufficient for practical key renewal, on a standard dark fiber (loss 0.25 dB/km) and for QKD systems operating without He-cooled detectors, under conservative security model (coherent attacks including finite-size effect, even if such analysis was not yet performed 10 years ago).
\end{thebibliography}
\end{document}